# The Greenland Telescope: Antenna Retrofit Status and Future Plans

Philippe Raffin[*c], Paul T.P. Ho[a], Keiichi Asada[a], Raymond Blundell[b], Geoffrey C. Bower[c], Roberto Burgos[b], Chih-Cheng Chang[a], Ming-Tang Chen[a], Robert Christensen[d], You-Hua Chu[a], Paul K. Grimes[b], C.C. Han[a], Chih-Wei L. Huang[a], Yau-De Huang[a], Fang-Chia Hsieh[d], Makoto Inoue[a], Patrick M. Koch[a], Derek Kubo[c], Steve Leiker[b], Lupin Lin[a], Ching-Tang Liu[d], Shih-Hsiang Lo[a], Pierre Martin-Cocher[a], Satoki Matsushita[a], Masanori Nakamura[a], Zheng Meyer-Zhao[a], Hiroaki Nishioka[a], Tim Norton[b], George Nystrom[a], Scott N. Paine[b], Nimesh A. Patel[b], Hung-Yi Pu[a], William Snow[c], T.K. Sridharan[b], Ranjani Srinivasan[c], and Jackie Wang[a]

[a] Academia Sinica Institute of Astronomy & Astrophysics, P.O. Box 23-141, Taipei 10617, Taiwan
[b] Harvard-Smithsonian Center for Astrophysics, 60 Garden Street, Cambridge, MA 02138, U.S.A.
[c] ASIAA Hilo Office, University Park, 645 N A'Ohoku Pl, Hilo, HI 96720, U.S.A.
[d] National Chung-Shan Institute of Science and Technology, N.300-5, Lane 277, Xi An Street, Xitun District, Taichung City 407, Taiwan

## ABSTRACT

Since the ALMA North America Prototype Antenna was awarded to the Smithsonian Astrophysical Observatory (SAO), SAO and the Academia Sinica Institute of Astronomy & Astrophysics (ASIAA) are working jointly to relocate the antenna to Greenland. This paper shows the status of the antenna retrofit and the work carried out after the recommissioning and subsequent disassembly of the antenna at the VLA has taken place.

The next coming months will see the start of the antenna reassembly at Thule Air Base. These activities are expected to last until the fall of 2017 when commissioning should take place. In parallel, design, fabrication and testing of the last components are taking place in Taiwan.

## 1. INTRODUCTION

Since the Antenna was awarded to the Smithsonian Astrophysical Observatory (SAO) in 2011, the project partners – ASIAA and SAO have been working towards retrofitting the antenna for cold weather operation prior to shipping to Greenland for millimeter and submillimeter VLBI and submillimeter single-dish observations. This 12-meter antenna – the Greenland Telescope (GLT) - will provide direct confirmation of a Super Massive Black Hole (SMBH) by observing its shadow image in the active galaxy M87 [4].

At the Montreal conference in 2014 we reported on antenna recommissioning made at the ALMA Test Facility in New Mexico, subsequent antenna disassembly, and major retrofitting tasks required enabling antenna operation in the extreme arctic environment of the Greenland site [1].

A project risk review, which was held at ASIAA in Taipei in April 2015, endorsed plans to deploy the antenna to Thule Air Base (TAB) for VLBI operation at 86 and 230 GHz prior to installing the telescope at Summit Station for higher frequency VLBI and submillimeter single-dish operation. The single-dish science case as well as up-to-date Summit submillimeter weather information can be found in [5]. Here we report on progress made towards antenna refurbishment.

## 2. ANTENNA RETROFIT

The following two tables show the antenna mechanical characteristics and site characteristics.

---

* Send correspondence to P. Raffin, e-mail: raffin@asiaa.sinica.edu.tw

*Antenna characteristics*

| | |
|---|---|
| D, Primary Aperture, meter | 12.000 |
| $f_p$, Primary Focal Length, meter | 4.800 |
| Primary $f_p$ / D ratio | 0.4 |
| Secondary Aperture, ds, meter | 0.750 |
| Final f / D ratio | 8.0 |
| M, Magnification Factor | 20.000 |
| $\theta_p$, Primary Angle of Illumination, degrees | 128.02 |
| $\theta_s$, Secondary Angle of Illumination, degrees | 7.16 |
| 2c, Distance between Primary and Secondary Foci, meter | 6.177 |
| H, Depth of Primary, meter | 1.875 |
| a, Distance from Elevation Axis to Focus, meter | 0.803 |
| y, Primary Vertex Hole Clear Aperture, meter | 0.750 |

Table 1: Antenna Optical Layout

*Site characteristics*

Table 2 indicates the site characteristics for which the antenna retrofit is taking place.

| | Primary operating conditions | Secondary operating conditions | Survival |
|---|---|---|---|
| Ambient temperature | 0 to -50 °C | 0 to -55°C | -73 °C |
| Vertical temperature gradient | 1 K.m$^{-1}$ | | n/a |
| Wind | 0 to 11 m.s$^{-1}$ | 0 to 13 m.s$^{-1}$ | 55 m.s$^{-1}$ |
| Winter Opacity, tau | Monthly Median 0.061, Monthly Min 0.049, Monthly Max 0.071 | | |
| Summer Opacity, tau | Monthly Median 0.134, Monthly Min 0.085, Monthly Max 0.189 | | |

Table 2: Characteristics of Summit Station, ultimate site for the GLT.

Opacity measurements at Summit Station taken with a 225 GHz tipping radiometer from August 2011 until April 2014 are described by Martin-Cocher who also shows a comparison of Summit Station with other submillimeter sites [3].

After the disassembly and testing of the ALMA NA prototype antenna in New Mexico at the VLA site, the antenna components had been separated into 3 types: to be re-used for the GLT without modification, to be re-used after modifications, and to be discarded and re-designed with new parts made. The 24 carbon fiber segments constituting the reflector backup structure (BUS), the elevation shafts and elevation bearings are the main components that were modified after disassembly at the VLA. These modifications have been described in detail in an earlier paper [1]. The support cone, azimuth bearing, cabin HVAC system, the whole servo system, all cables, quadrupod and hexapod are new components. In addition we designed a primary mirror panel de-ice system, side containers to house the servo cabinets and electronics, a platform to access the cabin via an arctic entrance. A single multiuse lifting spreader has been designed and fabricated for all antenna lifts and a steel foundation to support the antenna at TAB is also ready to ship to Greenland. These additional developments are presented here.

*Antenna side containers for electronics; platform*

The design has evolved since the previous conference, and the electronics, control and power components will reside in 2 side custom-designed containers as shown below. Figure 6 shows the 3D model of the GLT. The platform supporting the containers and the arctic access to the receiver cabin is the interface to the antenna mount and had to be entirely redesigned. Fabrication of these components is about to start in Taiwan. Integration of the antenna servo containers and associated electrical systems will take place at the Aeronautical Research Laboratories in Taichung. This laboratory, part of Taiwan National Chung-Shan Institute of Science and Technology has been closely working with us on the GLT

Project since the beginning. We plan to ship to Thule the equipped containers and platform with the July 2017 Thule Air Base Supply ship out of Norfolk, VA.

*Hexapod*

ADS International has delivered a new hexapod compliant with GLT requirements. Kinematically it is able to translate the subreflector in the axes orthogonal to the radio axis by ±5 mm while it is ±10 mm along the third degree of freedom and tilt it by ±1.5 degrees about the aperture's axes. The accuracy and repeatability in focus adjustment mode is better than 5 micrometers and each angular resolution can be commanded to better than 0.001°. For the chopping requirements, the range of 10-arc minute as well as the accuracy of the angular position of less than 1.3 arcsecond r.m.s. are met. The chopping frequency obtained is 1 Hz and the settling time 300 ms. The hexapod has been tested in an environmental chamber at the lowest operational temperature of -55 °C and survived at -70 °C.

*Panel de-icing system*

The purpose of the de-icing system is to raise the temperature of the aluminum panels by 2°C ±0.5 above ambient in order to prevent ice formation on the panel surface. A silver wire embedded in a silicon pad attached to the back of each panel provides the heating source. A solid piece of foam insulates the pad and prevents heat loss. The power supplied to the de-ice system is 10 kW. The surface is divided into 12 zones, each comprising 22 panels, and can be heated independently. We asked AIRBORNE, ALMA-NA manufacturer of the backup structure to drill additional holes in the segments for deicing cable routing, after verifying by finite element analysis that the structural integrity of the carbon fiber segments, the dish accuracy and pointing performances were not affected. The system is based on the South Pole Telescope design [7] as already explained earlier [1]. The system has been successfully tested in an environmental chamber down to -73 °C.

*Thule steel foundation*

The temporary site testing for the GLT operations in Thule is located at "Cluster Pad 1", a former airplane-parking pad close to the airfield. The ground is composed of an asphalt pavement on top of filled soil over the ice-rich permafrost covering the bedrock. The glacial sediment varies in average from 10 to 30 feet [6]. For stability during operations and to guarantee pointing accuracy, a steel foundation (Figure 1) has been designed which will be placed on wood pads to better distribute the load. In addition, the steel foundation and wood pads will be connected and the wood pads secured to the ground by a hold down mechanism to overcome the antenna's overturning moment under survival wind conditions. A grout will be used to fill the gaps between the load transfer interfaces between the foundation and cluster pad.

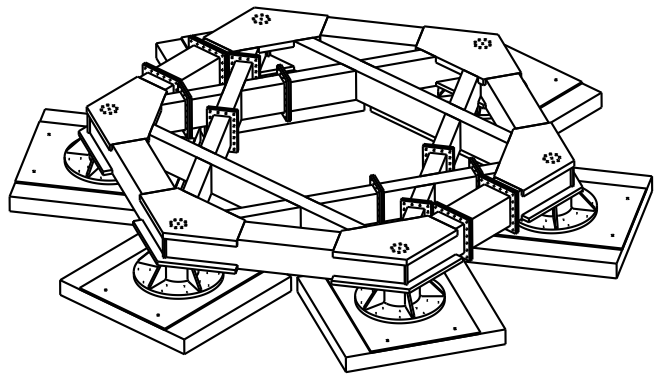

Figure 1: Steel foundation for the GLT operations at Thule Air Base.

*Lifting spreader*

We designed a multiuse lifting device, usable for lifting each component and each necessary assembly, including the assembled reflector. Figure 2 shows a 3D model of the lifting spreader in its stage to lift the support cone. To lift the traverse, the cabin/yoke arms or the invar cone, other lifting points are used. The right side of Figure 2 shows the spreader ready to lift the reflector backup structure and assembled quadrupod. The spreader has the capacity to lift the whole dish, including primary mirror panels, cladding panels, hexapod and subreflector.

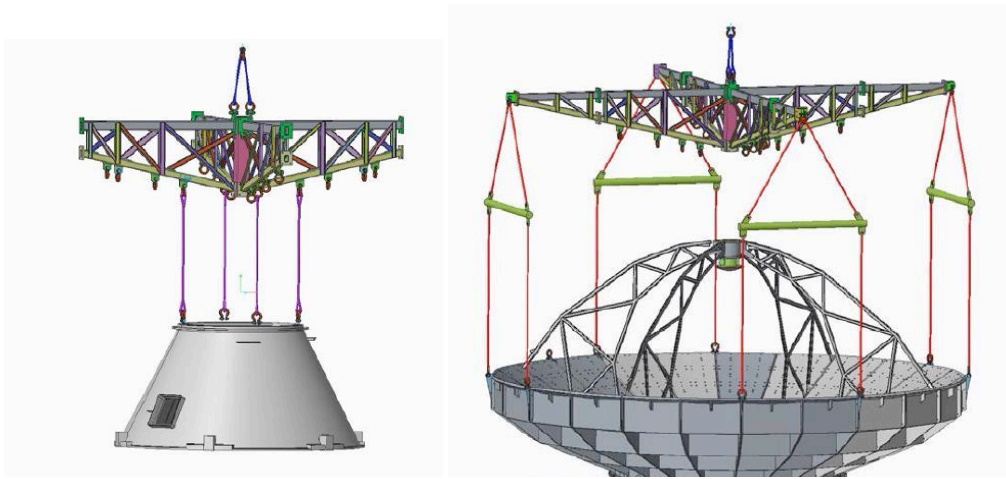

Figure 2: Left: lifting spreader for the support cone; Right: Lifting the 12-meter reflector with the extended spreader.

## 3. ANTENNA PRE-ASSEMBLY

### 3.1 Goals

The goals of the antenna pre-assembly were twofold: preassemble and install as many individual parts as achievable before shipping to Greenland, then disassemble into a few large subassemblies, of transportable size, with as much cabling, electronics, and insulation as possible left in place. While much of the antenna base was assembled at the Norfolk site, the assembly of some critical elements, such as the entire dish, including backup structure and reflector, will only take place at TAB. The GLT team has already been trained on backup structure assembly procedures at VERTEX Antennentechnik (VA), the prime contractor for this antenna, and the reflector panels can only be assembled once the dish is mounted on top of the antenna mount.

### 3.2 Achievements

We have secured a site close to the naval dockyard to store, assemble and test all but the main dish; and many of the critical components required to assemble the antenna up to the elevation axis were delivered. Assembly started early 2016 on the premises of World Distribution Services (WDS) of Norfolk, Virginia, and activities will conclude with shipment at the end of June 2016.

*Elevation part assembly and alignment of the axes*

To achieve this goal, we had to mount the complete antenna pedestal to be able to align the azimuth and elevation axes, as shown in Figure 3. The new support cone, already equipped with the new azimuth bearing, operation done at the factory in Germany, was first staged on steel plates on the ground. The 6 adjustable jacks interfacing the support cone to the foundation were used in conjunction with side hydraulic jacks. We mounted the yoke traverse on the azimuth bearing, and the yoke arms on the traverse as well as one of the 2 azimuth gearboxes and one motor. This allowed rotating the azimuth part with a temporary drive control box specially designed for this purpose by VA. The unit can drive one azimuth motor and one elevation motor. It is needed to rotate the antenna axes during the mechanical installation phase at Norfolk, but also in Thule, prior to electrical installation and commissioning of the GLT servo system. One motor at a time can be operated in jog control mode with adjustable velocity. Leveling of the azimuth bearing was repeated using the drive control box and an inclinometer temporarily attached to one of the yoke arms. Meanwhile, PSL engineers and technicians installed the elevation bearings while the receiver cabin was on the ground. Only at that point is the antenna ready to receive the elevation shafts. After assembly and alignment of the axes using a theodolite placed on the azimuth encoder support inside the support cone within the tolerances, the installation of the elevation bearings could be finalized. After measurement the cabin was off by 3.5 mm along elevation axis. We translated the cabin by this amount after removing the jacks temporarily supporting the cabin during installation of the elevation bearings. This was achieved by shimming along the shaft from the outside and then compensated on the other shaft to remove the gap. The azimuth axis verticality is 14.2", which is within the tolerable 20". The center point of the azimuth bearing to the vertical axis is 0.1 mm; the orthogonality between axes is 2 arcseconds. The center point of the elevation axis to the vertical axis is within 0.4 mm less than 0.5 mm being the maximum misalignment permitted. Figure 3 shows the antenna mount after axes alignment.

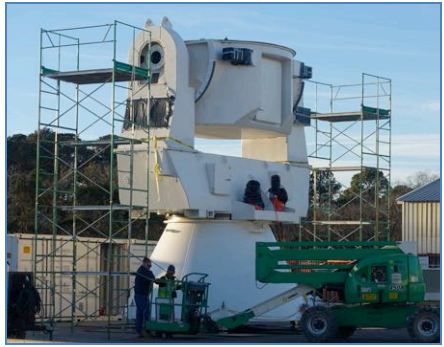

Figure 3: Elevation bearings installed, cabin mounted and aligned

*HVAC installation and testing*

VA's thermal consultant, the Institute of Energy, Cooling and Air Conditioning Technology did the study of the HVAC system and insulation. The dimensioning was based on winter and summer worst-case scenarios at Summit Station. In winter for the worst primary operating mode, the outside air temperature is -55°C with a 13 m/s wind, no sun while the receiver cabin inside temperature stays at 22°C. The other worst-case scenario is in survival mode with an outside air temperature of -73°C and a 55 m/s wind for the same receiver cabin temperature. This leads to heat losses of respectively 3.5 and 4.2 kW, which have to be recovered. In summer for the worst thermal load case we will also have a slight heat loss from the receiver cabin to the ambient, subject to similar winds, but with an asymmetric irradiation of 800 W/m$^2$ of the cabin, combined with the radiation of the snow on the other side. A total cooling capacity of about 4 kW is needed. An outside unit to be placed under the antenna platform will provide this amount. In addition, to maintain the temperature homogeneity in the receiver cabin from +15 to +22°C ±1°C and to guarantee pointing accuracy, a cabin wall tempering system is necessary as well as an elevation bearing and shaft tempering system. A network of copper pipes with flowing coolant is running vertically along the walls of the receiver cabin and along the elevation shafts. While the cabin was on the floor and prior to bearing installation, the complete tempering system was installed by MENERGA, and tested for leaks. The 2 inside racks recirculating and filtering the air and the glycol distribution unit were installed. It was not possible to test the whole system in Norfolk and this will be done at TAB.

*Electrical outfit*

Cable installation at Norfolk was limited and took place only inside the cabin, yoke and support cone: installation of electronic bus stations, terminal boxes, temperature sensor holders and wiring to next box; installation of e-stops and cabling to local components, and installation of lighting and sockets. Cabling between antenna components can only take place at TAB and will use low-temperature cables and connectors.

*Antenna insulation*

The thermal study also recommended covering all parts of the antenna mount with a high-quality thermal insulation. A 3-inch thick elastomer foam for cryogenic use based on composition rubber, guarantying flexibility down to -196 °C and providing a thermal conductivity of $\lambda$ = 0.034 W/m×K at -50°C and 0.028 W/m×K at -100°C. A sealed aluminum sheet preventing moisture ingress at the final site will protect the foam. Most of the antenna components have been insulated with foam and aluminum sheeting in Norfolk. The insulation will be finalized in Greenland. Unlike other antennas for which insulation is the final step in the assembly process, the difficulty of this task is increased, as we need to provide access to bolts areas to dismount the antenna in subcomponents for shipping to Thule, and later to Summit Station. The insulation work at Norfolk was only possible because we had tents installed around antenna structural components as shown below, given the rainy winter and spring of Eastern Virginia.

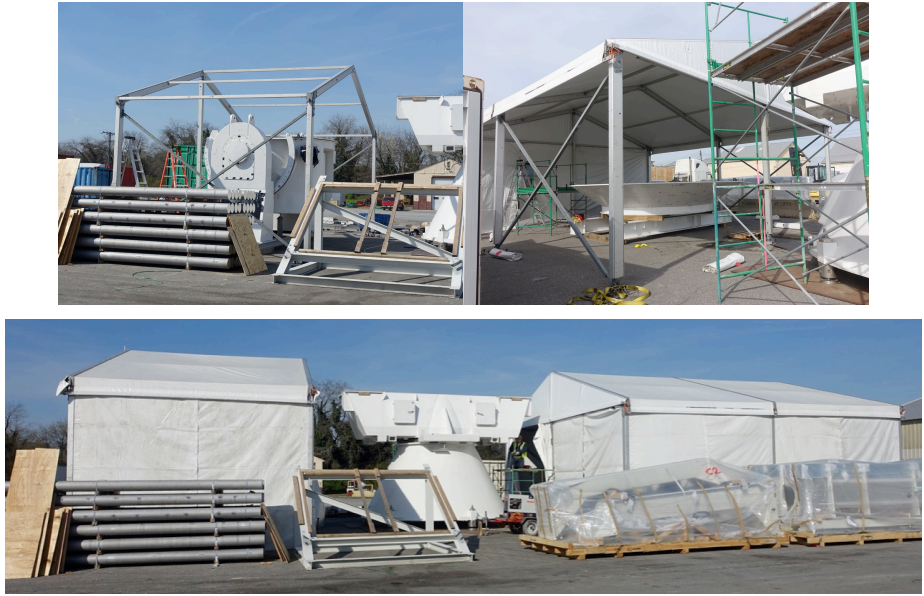

Figure 4: Tents necessary for antenna insulation work at Norfolk during winter and spring 2016.

*Trailer outfit*

Three standard 40-ft containers have also been transformed into work trailers and outfitted for cold weather operation in Greenland, they are: mechanical workshop, control room and electrical laboratory. The control trailer will house the control and monitoring computers and equipment to operate the antenna. It will also contain all the precision timing devices (Maser, frequency referencing device, GPS, etc.) and the first lag of primary data storage off the antenna.

## 4. PREPARATION FOR SHIPPING

*Antenna disassembly*

The antenna has been separated into 4 out-of-gauge (OOG) components for shipping in compliance with cargo requirements in size and weight: the support cone and the azimuth bearing, the yoke traverse, the yoke arms and receiver cabin, and the invar cone. All other components are collected from the various sites around the world where they have been retrofitted or built anew to WDS. From there containers will be trucked to the supply ship at Navy Pier 8, while out-of-gauge parts will first be trucked to an intermediate location then transferred to the supply ship by barge. The supply ship is expected to leave early July from the U.S. Navy Base at Norfolk and arrive at TAB mid-July.

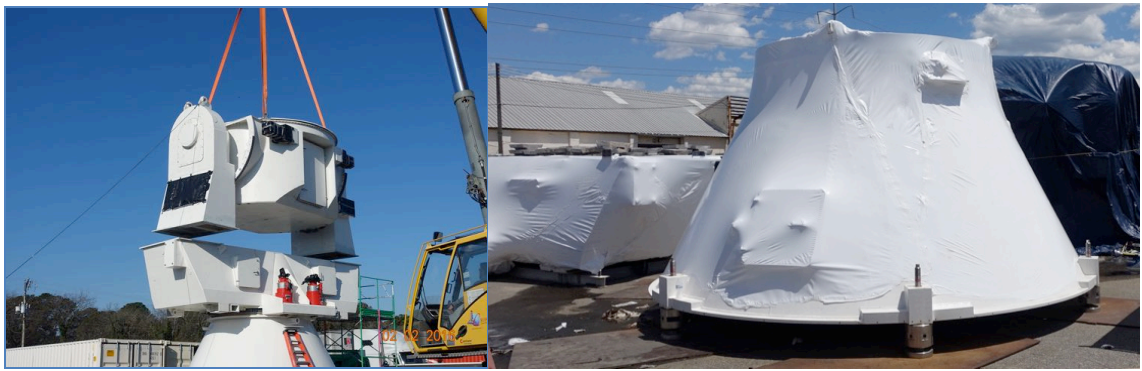

Figure 5: Left: Disassembly of the receiver cabin/yoke arms unit from the yoke traverse, before insulation work; Right: Shrink-wrap of the support cone with the azimuth bearing and lifting fixture integrated is in progress.

*Packing*

We plan to shrink wrap all OOG components at WDS for the ocean journey as shown in above picture on the right hand side of Figure 5. The components are planned to be stored below deck. All but the quadrupod crate holding the 4 quadlegs disassembled, fit in regular size containers.

## 5. FUTURE PLANS

*Plans for Thule*

Working with the US National Science Foundation, we have identified a site for antenna deployment at TAB; and negotiations have taken place with Base Command to enable assembly to proceed as soon as the antenna components arrive in Greenland in July 2016. The first assembly phase is expected to last until spring of 2017. It will comprise staging the steel foundation at the site, reassembly of the antenna mount; finalize servo components installation, antenna insulation, and perform basic antenna tests. Control and monitoring software is being developed and a version is ready for the initial tests of the antenna at Thule as presented at this conference by Patel [2]. Optical and radio pointing calibration, holography, and VLBI observations at 230 GHz will be carried out in 2017. In parallel, the reflector backup structure will be assembled in a hangar of the base and 800 new panel adjusters will replace one type of old adjusters, not compatible with low-temperature operation at Summit Station. The second assembly phase will start in the spring of 2017 and will see the backup structure moved out of the hangar and installed on the antenna. Panel assembly will then be done, but precise alignment and photogrammetry will only be done later, once the HVAC and side containers are installed. The receiver cabin needs to be at the operating temperature, which will happen only during the final assembly phase. The third assembly phase will start in summer of 2017 when the last bulk of components are ready and shipped to TAB, mainly the antenna side containers and platform, the arctic entrance and the cables. We plan to completely integrate the electronic components in the containers and do the internal wiring in Taiwan, before shipping to Greenland. These containers will be mounted on the antenna at TAB; cables installed and final testing will take place for commissioning. The telescope commissioning should continue through summer and fall 2017, and early science tests are expected to start during the winter 2017-18.

*Plans for Summit Station*

We plan to disassemble the GLT antenna into large subcomponents of tractable size for the Greenland Inland Traverse (GrIT) [8], which will bring the antenna parts and all other necessary equipment to Summit Station after a 1,000 km journey and reassemble the antenna during the same year at the GLT final site. This may be achievable as early as 2019. The journey takes about 4 weeks and can only start early spring when access to the ice sheet is possible; this is dictated by snow condition and daylight. Prior to the assembly at Summit Station, the snow pad, serving as foundation for the antenna will be prepared. The snow pad has to settle for one year, before we can install the space frame. The goal of the space frame is to distribute the load of the telescope to the foundation. The outer area around the foundation will be shaped to delay snow accumulation and drifts around the telescope. This structure has already been described earlier [1].

## REFERENCES

[1] Raffin, P., et al. "The Greenland Telescope (GLT): Antenna status and future plans", Proc. SPIE Montreal, Vol. 9145-15 (2014)
[2] Patel, N. A., "Control and monitoring software for the Greenland Telescope", SPIE Edinburgh (2016)
[3] Martin-Cocher, P. L., et al. "225-GHz opacity measurements at Summit camp, Greenland, for the Greenland telescope (GLT) site testing", SPIE Montreal, Vol. 9147-138 (2014)
[4] Inoue et al. 2014, Radio Science, 49, 564
[5] Hirashita et al. 2016, PASJ, 68, R1
[6] Bjella, Kevin, "Thule Air Base Airfield White Painting and Permafrost Investigation", Final report approved for public release, Cold Regions Research and Engineering Laboratory U.S. Army Engineer Research and Development Center 72 Lyme Road Hanover, NH 03755, (2013)
[7] Carlstrom, J. E. et al. "The 10 Meter South Pole Telescope", PASP, 123, 568 (2011)

[8] Lever, J.H. "Greenland Inland Traverse", National Science Foundation, United States Antarctic Program (USAP) and Arctic Sciences Program (ARC), Cold Regions Research and Engineering Laboratory U.S. Army Engineer Research and Development Center (2011).

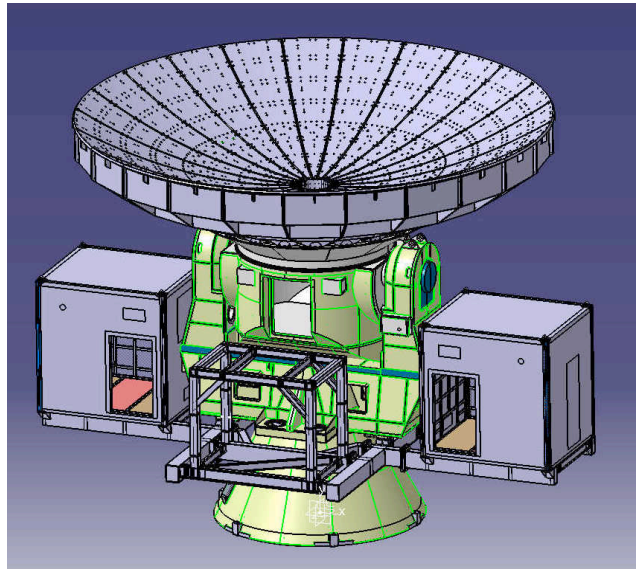

Figure 6: GLT antenna 3D-model showing the 2 servo and electronics side containers.